\documentclass[seceq]{iis}
\usepackage{graphics}
\usepackage{color}
\usepackage{amsmath}
\usepackage{setspace}
\usepackage{caption}

\title{
A useful way to obtain the central charge of entanglement Hamiltonian\\
-- Nested entanglement entropy --
}
\runtitle{
Nested entanglement entropy
}

\author{%
\name{Shu \surname{TANAKA}}
\CAE{shu-t@chem.s.u-tokyo.ac.jp}, 
}
\inst{
Department of Chemistry, University of Tokyo, \address{Hongo, Bunkyo-ku, Tokyo 113-0033, Japan}
}
\runauthor{Shu Tanaka}

\support{
During the studies related to this manuscript, the author was supported by Grant-in-Aid for Young Scientists start-up (21840021) from the JSPS and JSPS Fellows (23-7601) and numerical calculations were performed on supercomputers at the Institute for Solid State Physics, University of Tokyo.
}

\recdate{20YY}{0}{DD}%% set by the editors
\accdate{20YY}{0}{DD}%% set by the editors

\abst{
In this paper, we review how to obtain the central charge of a critical entanglement Hamiltonian through the nested entanglement entropy which was first introduced in Ref.~\cite{Lou-2011}.
The critical phenomena of the entanglement Hamiltonian can be identified by the central charge obtained by the nested entanglement entropy.
We review our previous studies~\cite{Lou-2011,Tanaka-2012} in which we investigated certain entanglement nature of two-dimensional valence-bond-solid (VBS) state and quantum hard-square models on square and triangle ladders using the nested entanglement entropy.
}

\kword{\kw{Nested entanglement entropy}, \kw{Entanglement Hamiltonian}, \kw{Central charge}, \kw{Two-dimensional quantum systems}}

\begin{document}
\maketitle

\section{Introduction}
\label{Introduction}

Recently, the study of entanglement in quantum many-body systems has been done in a wide area of science {\it e.g.}, quantum information science, condensed matter physics, and quantum statistical physics\cite{Horodecki-2009,Eisert-2010}.
A useful measure of entanglement is the von Neumann entanglement entropy which is often used to quantify the degree of entanglement of a bipartite system consisting of two subsystems.
For example, one-dimensional quantum systems can be characterised by the entanglement entropy. 
In one-dimensional gapped systems, the entanglement entropy is bounded below by the logarithm of the number of edge states in the thermodynamic limit.
One of the most visible examples is shown in Ref.~\cite{Katsura-2007}, where the entanglement entropy of the valence-bond-solid (VBS) states with generic integer spin were studied.
The authors found that the saturation value is consistent with the edge-state picture.
In one-dimensional critical systems, on the other hand, the entanglement entropy diverges logarithmically with the chain length\cite{Vidal-2003}.
The coefficient of the logarithmic function relates to the central charge $c$.
In fact, the behavior of the entanglement entropy is similar to that of entropy in conformal field theory (CFT).
In CFT, the central charge classifies the universality class in two-dimensional critical phenomena.

Li and Haldane introduced entanglement spectrum which is the eigenvalue spectrum of the entanglement Hamiltonian generated from the reduced density matrix\cite{Li-2008}.
Compared to the entanglement entropy, the entanglement spectrum has more complete information on the ground state of the system.
The entanglement spectrum has been applied to many kinds of systems {\it e.g.}, topologically ordered systems and quantum spin models.
However, the application of the entanglement spectrum has been rather limited to one-dimensional or topological systems.
Very recently, the entanglement spectrum of two-dimensional quantum systems has been studied\cite{Yao-2010,Cirac-2011,Huang-2011,Lou-2011,Tanaka-2012}.
Cirac {\it et al.} and Lou {\it et al.} independently studied the entanglement spectra of two-dimensional valence-bond-solid (VBS) state\cite{Cirac-2011,Lou-2011} by different approaches.
The low-energy parts of the entanglement spectra of the VBS state on square and hexagonal lattices are respectively similar with the energy dispersions of antiferromagnetic and ferromagnetic Heisenberg spin chains.
Tanaka {\it et al.} considered the entanglement properties of the ground state of the quantum hard-square model\cite{Tanaka-2012}.
They found that the entanglement spectra of the ground state on square and triangle ladders are similar with the energy dispersions of the two-dimensional critical Ising model and the two-dimensional three-state Potts model, respectively.

If the entanglement spectrum is obtained, we can identify the central charge by comparing the obtained spectrum and the energy dispersion of a quantum critical system.
However, since the size effect is not negligible in some cases, development of a new method which provides the central charge has been needed.
To this end, we introduced a new quantity -- nested entanglement entropy which allows us to obtain the central charge of the entanglement Hamiltonian rather directly.
Using the nested entanglement entropy, we succeed to obtain the central charges of entanglement Hamiltonians generated from the VBS state on square lattice and the ground state of the quantum hard-square model on square and triangle ladders. 
In this paper, we review the nested entanglement entropy and the obtained results in our previous studies\cite{Katsura-2010,Lou-2011,Tanaka-2012}.

\section{Entanglement properties}

In this section, we review some basic properties of entanglement.
For simplicity, suppose we consider a bipartition of a system into two subsystems A and B, each of which is a mirror image of the other.
In other words, the boundary between A and B are chosen to be the reflection axis of the system.
Here we follow the approach by Shi {\it et al.} in Ref.~\cite{Shi-2006}.
Let us consider the case where a unnormalized ground state $|\psi\rangle$ of the total system is described by
\begin{eqnarray}
 \label{eq:state_AB}
 |\psi\rangle = \sum_\alpha 
  |\phi_\alpha^{[{\rm A}]} \rangle \otimes 
  |\phi_\alpha^{[{\rm B}]} \rangle,
\end{eqnarray}
where the superscripts [A] and [B] indicate the subsystems A and B, respectively.
The set of states $\{|\phi_\alpha^{[{\rm A}]}\rangle\}$ and $\{|\phi_\alpha^{[{\rm B}]}\rangle\}$ are linearly independent but may not be orthogonal.
To Schmidt decompose Eq.~(\ref{eq:state_AB}), we define the matrices as follows:
\begin{eqnarray}
 (M^{[{\rm A}]})_{\alpha\beta} := \langle \phi_\beta^{[{\rm A}]} | \phi_\alpha^{[{\rm A}]} \rangle,
  \quad
 (M^{[{\rm B}]})_{\alpha\beta} := \langle \phi_\beta^{[{\rm B}]} | \phi_\alpha^{[{\rm B}]} \rangle.
\end{eqnarray}
The spectral decompositions of the matrices are given by
\begin{eqnarray}
 M^{[{\rm A}]} = XD^{[{\rm A}]}X^\dagger,
  \quad
 M^{[{\rm B}]} = YD^{[{\rm B}]}Y^\dagger,
\end{eqnarray}
where $X$ and $Y$ are unitary matrices.
$D^{[{\rm A}]}$ and $D^{[{\rm B}]}$ are diagonal matrices where $(D^{[a]})_{\tau\tau'}=\delta_{\tau\tau'}d_\tau^{[a]}$ ($a=$A, B).
Orthogonal bases in the subsystems A and B can be obtained by using $X$ and $Y$:
\begin{eqnarray}
 |e_\tau\rangle = \frac{1}{\sqrt{d_\tau^{[{\rm A}]}}} \sum_\alpha (X^\dagger)_{\tau\alpha}|\phi_\alpha^{[{\rm A}]}\rangle,
  \quad
 |f_\eta\rangle = \frac{1}{\sqrt{d_\eta^{[{\rm B}]}}} \sum_\alpha (Y^\dagger)_{\eta\alpha}|\phi_\alpha^{[{\rm B}]}\rangle.
\end{eqnarray}
Then, the unnormalized ground state $|\psi\rangle$ is expressed as
\begin{eqnarray}
 |\psi\rangle = \sum_\tau\sum_\eta \sqrt{d_\tau^{[{\rm A}]} d_\eta^{[{\rm B}]}}
  (X^{\rm T}Y)_{\tau\eta}|e_\tau\rangle\otimes|f_\eta\rangle,
\end{eqnarray}
where ${\rm T}$ denotes matrix transpose.
Since we now consider the case where the subsystem A is a mirror image of the subsystem B, $M^{[{\rm A}]}=M^{[{\rm B}]}=M$.
If $M$ is a real symmetric matrix, $M$ can be diagonalized by an orthogonal matrix $O$: $M=ODO^{\rm T}$, where $(D)_{\tau\tau'}=\delta_{\tau\tau'}d_\tau$ where $d_\tau = d_\tau^{[{\rm A}]} = d_\tau^{[{\rm B}]}$.
Then, the unnormalized ground state Eq.~(\ref{eq:state_AB}) can be expressed as
\begin{eqnarray}
 \label{eq:state_decomposition}
 |\psi\rangle = \sum_\tau d_\tau |e_\tau\rangle \otimes |f_\tau\rangle.
\end{eqnarray}
Here we consider the density matrix of the total system defined by
\begin{eqnarray}
 \rho_{\rm tot} := \frac{|\psi\rangle\langle \psi|}{\langle \psi|\psi\rangle}.
\end{eqnarray}
The reduced density matrix for the subsystem A is given by
\begin{eqnarray}
 \rho_{\rm A} = {\rm Tr}_{\rm B} \rho_{\rm tot}
  = \sum_\eta \langle f_\eta | \rho_{\rm tot} |f_\eta\rangle = 
  \frac{\sum_\tau d_\tau^2 |e_\tau\rangle\langle e_\tau|}{\sum_\tau d_\tau^2}.
\end{eqnarray}
Then the von Neumann entanglement entropy is obtained by 
\begin{eqnarray}
 S = - {\rm Tr}\rho_{\rm A} \ln \rho_{\rm A} = - \sum_\tau p_\tau \ln p_\tau,
\quad
  p_\tau = \frac{d_\tau^2}{\sum_\tau d_\tau^2}.
\end{eqnarray}

\begin{figure}[t]
\centering
\includegraphics[scale=0.9]{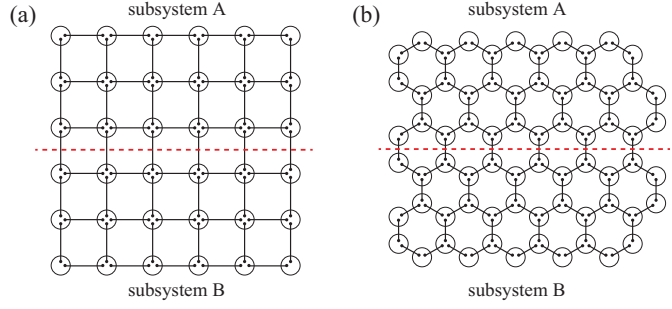}
\caption{
 (a) VBS state on square lattice. 
 (b) VBS state on hexagonal lattice.
 The red dotted lines indicate the boundaries.
 In each case, the periodic boundary condition is imposed in the boundary direction.
}
\label{fig:vbs}
\end{figure}

Next we explain the entanglement spectrum which was introduced by Li and Haldane\cite{Li-2008}.
The entanglement entropy is defined by
\begin{eqnarray}
 H_{\rm E}:= -\ln \rho_{\rm A}.
\end{eqnarray}
The entanglement spectrum is the energy dispersion of the entanglement Hamiltonian.
This provides more detailed information on the entanglement nature in comparison with the entanglement entropy.
When the eigenenergies of the excited states are separated from the eigenenergy of the ground state, the entanglement is weak, {\it i.e.}, the von Neumann entanglement entropy is small.
On the other hand, when all eigenenergies are degenerated, the state under consideration is a maximally entangled state.
Moreover, the entanglement spectrum provides an interesting nature of the entanglement Hamiltonian.
As will be explained later, the low-energy part of the entanglement spectrum is similar with the energy dispersion of one-dimensional quantum systems.
Lou {\it et al.} studied entanglement properties of two-dimensional VBS state on square and hexagonal lattices as shown in Fig.~\ref{fig:vbs}\cite{Lou-2011}.
The periodic boundary condition is imposed in the boundary direction.
Using the Schwinger boson representation, the VBS state can be written in the form Eq.~(\ref{eq:state_decomposition}) where $\tau=1,2,\cdots,2^{L_x}$ ($L_x$ is the number of edge sites along the boundary in the subsystem A).
The reduced density matrix is a $2^{L_x}\times 2^{L_x}$ matrix and positive definite.
The entanglement entropy per boundary site converges to a certain value which is less than $\ln 2$.
The low-energy part of the entanglement spectrum of the VBS state on square (resp. hexagonal) lattice resembles the energy dispersion of the spin-$1/2$ antiferromagnetic (resp. ferromagnetic) Heisenberg chain.

\begin{figure}[b]
\centering
\includegraphics[scale=0.9]{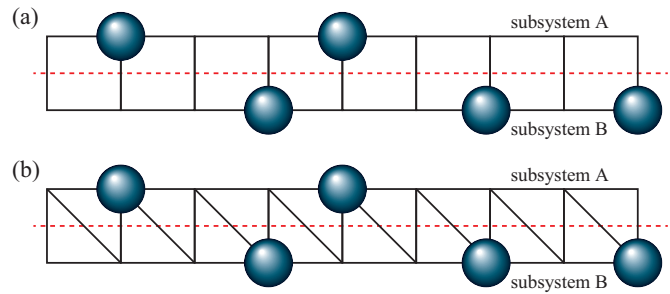}
\caption{
 (a) Quantum lattice gas model on square ladder. 
 (b) Quantum lattice gas model on triangle ladder.
 The red dotted lines indicate the boundaries.
 In each case, the periodic boundary condition is imposed in the boundary direction.
 }
\label{fig:qhsm}
\end{figure}

Tanaka {\it et al.} considered certain entanglement properties of quantum lattice gas model on square and triangule ladders as shown in Fig.~\ref{fig:qhsm}\cite{Tanaka-2012}.
The periodic boundary condition is imposed in the boundary direction as in the case of the VBS state.
In order to ``make'' a Hamiltonian in which the ground state is obtained exactly, the following two conditions are imposed.
The one is that there is at most one particle at each site, {\it i.e.}, hard-core boson.
The other is that there is no more than one boson on any pair of nearest-neighbor sites.
Because of the first condition, the Hilbert space at each site is spanned by $|n_i\rangle$, where $n_i=0 ({\rm resp.}\, 1)$ means that the site $i$ is empty (resp. occupied).
With the identification, $|0\rangle \leftrightarrow |\downarrow\rangle$ and $|1\rangle \leftrightarrow |\uparrow\rangle$, the Hamiltonian of the system is described by 
\begin{eqnarray}
 \label{eq:model-qhsm}
 H = \sum_i h_i^\dagger (z) h_i (z),
  \quad
  h_i(z) = [\sigma_i^- - \sqrt{z}(1-n_i)] P_{\langle i \rangle},
\end{eqnarray}
where $z$ is the fugacity, $\sigma_i^-$ is defined by $\sigma_i^- = (\sigma_i^x - i \sigma_i^y)/2$ ($\sigma_i^\alpha$: Pauli matrices), $n_i$ is the number operator, and $P_{\langle i \rangle}$ is the projection operator which is introduced so as to satisfy the second condition.
Since the Hamiltonian is positive semidefinite, a zero-energy state is a ground state. It then follows from the Perron-Frobenius theorem that this state is the unique ground state of $H$.
The unnormalized ground state is described by the weighted superposition of classical configurations:
\begin{eqnarray}
 \label{eq:gs_qhsm}
 |\psi(z)\rangle = \sum_{C \in S} z^{n_C/2}|C\rangle,
\end{eqnarray}
where $C$ label classical configurations of particles on the lattice and $S$ is the set of configurations with nearest-neighbor exclusion.
The ground state can be described by Eq.~(\ref{eq:state_decomposition}) where $\tau=1,2,\cdots,L$ ($L$ represents the length of the ladder).
Then, a treatment similar to that used in the case of the VBS state.
Since the ground state at $z=0$ is the vacuum state, the entanglement entropy is zero.
The entanglement entropy in the limit of $z\to \infty$ for square (resp. triangle) ladder converges to $\ln 2$ (resp. $\ln 3$), which is consistent with the edge-state picture.
The authors obtained the entanglement spectrum of the ground state at the critical fugacity $z_{\rm c}$ of the corresponding classical systems on square and triangle lattices.
The entanglement spectrum for square (resp. triangle) ladder is reminiscent of energy dispersion of the two-dimensional critical Ising model (resp. the two-dimensional three-state Potts model).
%%%

\section{Nested entanglement entropy}

As shown in the previous section, the relation between the holographic system and the one-dimensional quantum system is observed.
In order to further establish the relation, we first introduced a measure called nested entanglement entropy\cite{Lou-2011}.
The nested entanglement entropy can be calculated using only the ground state of $H_{\rm E}$.
For one-dimensional critical systems, the underlying CFT can be read off from the von Neumann entanglement entropy of the ground state.
Thus, we can directly obtain the central charge of the entanglement Hamiltonian from the scaling analysis of the nested entanglement entropy.

\begin{figure}[h]
\centering
\includegraphics[scale=0.6]{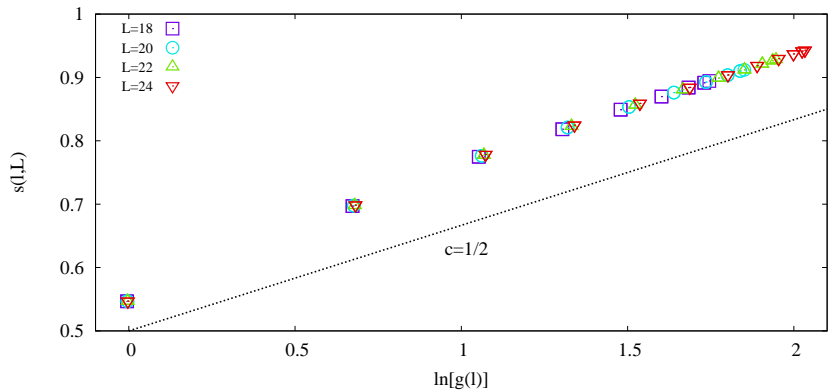}
\caption{
 The nested entanglement entropy of the quantum lattice gas model on square ladder with $L=18,20,22,24$.
 The dotted line indicates $c=1/2$.
 }
\label{fig:nee}
\end{figure}

Suppose we divide the system described by the entanglement Hamiltonian with the length $L$ into two subsystems.
The length of one subsystem is $\ell$ and that of the other is $L-\ell$.
The definition of the nested density matrix is 
\begin{eqnarray}
 \rho(\ell) := {\rm Tr}_{\ell+1,\cdots,L}[|\phi_0\rangle\langle \phi_0|],
\end{eqnarray}
where $|\phi_0\rangle$ is the ground state of $H_{\rm E}$ and ${\rm Tr}_{\ell+1,\cdots,L}$ means the trace out over the degree of freedom of the subsystem with the length $L-\ell$.
The nested entanglement entropy is defined by
\begin{eqnarray}
 s(\ell,L):= -{\rm Tr}_{1,\cdots,\ell} [\rho(\ell)\ln\rho(\ell)],
\end{eqnarray}
where the trace is taken over the states in the subsystem with the length $\ell$.
At the ``critical'' point, the nested entanglement entropy for systems with periodic boundary condition is expected to be as
\begin{eqnarray}
 \label{eq:nee_relation}
 s(\ell,L)= \frac{c}{3}\ln [g(\ell)]+s_1,
  \quad
g(\ell) = \frac{L}{\pi}\sin\left(\frac{\pi \ell}{L}\right),
\end{eqnarray}
where $c$ is the central charges of the entanglement Hamiltonian and $s_1$ is a non-universal constant\cite{Calabrese-2004}.
Figure \ref{fig:nee} shows the nested entanglement entropy of the quantum lattice gas model on square ladder.
The gradient of the dotted line indicates $c=1/2$.
Then, the behavior of the nested entanglement entropy concludes that the central charge of this case is $c=1/2$.
By using the relation given in Eq.~(\ref{eq:nee_relation}), the central charge of certain entanglement Hamiltonians has been determined, as summarized in Table \ref{table:cft}.
The central charge obtained by the nested entanglement entropy establishes the critical properties of the entanglement Hamiltonian rather than the entanglement spectrum.
It should be noted that the entanglement Hamiltonian is critical, although the original system is gapped.
An intriguing fact emerges through the analysis of entanglement properties.

\begin{table}[h]
 \caption{Correspondence between the physical system and the CFT of the entanglement Hamiltonian.}
 \label{table:cft}
 \begin{center}
 \begin{tabular}[t]{c|c}
  \hline
  Physical system & Central charge of the entanglement Hamiltonian \\
  \hline
  VBS state on square lattice\cite{Lou-2011} & $c=1$ (spin-$1/2$ antiferromagnetic Heisenberg chain)\\
  Ground state of the hard-square model on square ladder at $z=z_{\rm c}$\cite{Tanaka-2012} & $c=1/2$ (Critical Ising model in two dimension)\\
  Ground state of the hard-square model on triangle ladder at $z=z_{\rm c}$\cite{Tanaka-2012} & $c=4/5$ (Three-state Potts model in two dimension)\\
  \hline
 \end{tabular}
 \end{center}
\end{table}

\section*{Acknowledgments}

The author acknowledges the collaboration with Hosho Katsura, Naoki Kawashima, Vladimir E. Korepin, Anatol N. Kirillov, Jie Lou, and Ryo Tamura, on which this manuscript is based.
The author wishes to make my deep acknowledgement to Hosho Katsura and Ryo Tamura for their careful reading of the manuscript.
The author thanks Shunsuke Furukawa, Toshiya Hikihara, Yutaka Shikano, and Takafumi Suzuki for the fruitful discussions.

\end{document}